\documentclass[12pt]{iopart}
\pdfoutput=1

\expandafter\let\csname equation*\endcsname\relax
\expandafter\let\csname endequation*\endcsname\relax

\usepackage{graphicx}
\usepackage{amsmath} 
\usepackage{amssymb}
\usepackage{xcolor}
\usepackage[colorlinks=true,allcolors=black]{hyperref}
\usepackage{subfig}
\usepackage{times}

\newcommand{\ket}[1]{|#1 \rangle}

\begin{document}
\graphicspath{{pictures/}}
\title{Laser threshold magnetometry}

\author{Jan Jeske$^1$, Jared H.~Cole$^1$, Andrew D.~Greentree$^{1,2}$}
\address{$^1$Chemical and Quantum Physics, School of Applied Sciences, RMIT University, Melbourne 3001, Australia\\
$^2$Australian Research Council Centre of Excellence for Nanoscale BioPhotonics}
\ead{janjeske@gmail.com}

%
%


\begin{abstract}
We propose a new type of sensor, which uses diamond containing the optically active nitrogen-vacancy (NV$^-$) centres as a laser medium. The magnetometer can be operated at room-temperature and generates light that can be readily fibre coupled, thereby permitting use in industrial applications and remote sensing. By combining laser pumping with a radio-frequency Rabi-drive field, an external magnetic field changes the fluorescence of the NV$^-$ centres. We use this change in fluorescence level to push the laser above threshold, turning it on with an intensity controlled by the external magnetic field, which provides a coherent amplification of the readout signal with very high contrast. This mechanism is qualitatively different from conventional NV$^-$--based magnetometers which use fluorescence measurements, based on incoherent photon emission. We term our approach laser threshold magnetometry (LTM). We predict that an NV$^-$--based laser threshold magnetometer with a volume of 1mm$^3$ can achieve shot-noise limited d.c.~sensitivity of 1.86 fT$/\sqrt{\rm{Hz}}$ and a.c.~sensitivity of 3.97fT$/\sqrt{\rm{Hz}}$.
\end{abstract}


\maketitle
The precise measurement of magnetic fields (magnetometry) has a variety of scientific applications including NMR detection and gravitational wave detection. More generally, it is an enabling technology for mining exploration, to detect oil, gas and mineral reserves \cite{Nabighian2005, Foley2002}, in airports for automated detection of plane movements with less building-induced interference than radar \cite{Tamayo1993, Stockhammer2006}, and in medicine for the detection of magnetic fields produced by the heart (magneto-cardiography - MCG) or brain (magneto-encephalography - MEG) \cite{Hamalainen1993, Rodriguez1999, Gratta2001, Proudfoot2014}. New room-temperature operated sensors with better sensitivities would enable new mapping techniques at higher spatial resolution. Current state-of-the-art sensors with sensitivities around 1 fT/$\sqrt{\mathrm{Hz}}$ are the widely used SQUID magnetometers \cite{Weinstock1991, Drung2001}, operated at cryogenic temperatures below 10 K, and the more recent spin-exchange relaxation-free (SERF) atomic magnetometers \cite{Kominis2003}. Diamond containing negatively-charged nitrogen-vacancy (NV$^-$) centres has emerged as being important for niche applications including room-temperature nanoscale magnetometry \cite{Degen2008, Balasubramanian2008, Taylor2008, Maze2008}. 

The NV$^-$ colour centre in diamond possesses numerous outstanding properties, that have made it ideal for quantum applications.  In particular, because of efficient optical spin polarisation and readout \cite{Jelezko2004}, it has been identified as a biocompatible nanoscale magnetometer \cite{Balasubramanian2008, Taylor2008, Maze2008, McGuinness2011}, electrometer \cite{Dolde2011, Dolde2014}, thermometer \cite{Neumann2013, Kucsko2013, Plakhotnik2014} and quantum environment sensor \cite{Coledecoherencemicroscopy, Hall2009, Kaufmann2013}. For a comprehensive description of the NV$^-$ centre, see Doherty \textit{et al.} \cite{Doherty2013}.

Much of the work on NV$^-$ sensing has concentrated on the use of single centres, so as to achieve the smallest volume possible, however there have also been investigations of NV$^-$ ensemble-based magnetometry \cite{Charnock2001, Acosta2009, Aharonovich2009, Greentree2009, Pham2011}. Whilst such approaches provide improved signal-to-noise over single centres by increasing the number of atoms, such improvements are not predicted to give more than a $\sqrt{N}$ improvement, where $N$ is the total number of centres capable of sensing. In recent ensemble experiments new sensitivity records around pT/$\sqrt{\rm{Hz}}$ were achieved\cite{Clevenson2015, Wolf2014}.

\begin{figure*}
\subfloat[]{\includegraphics[width=0.35\columnwidth]{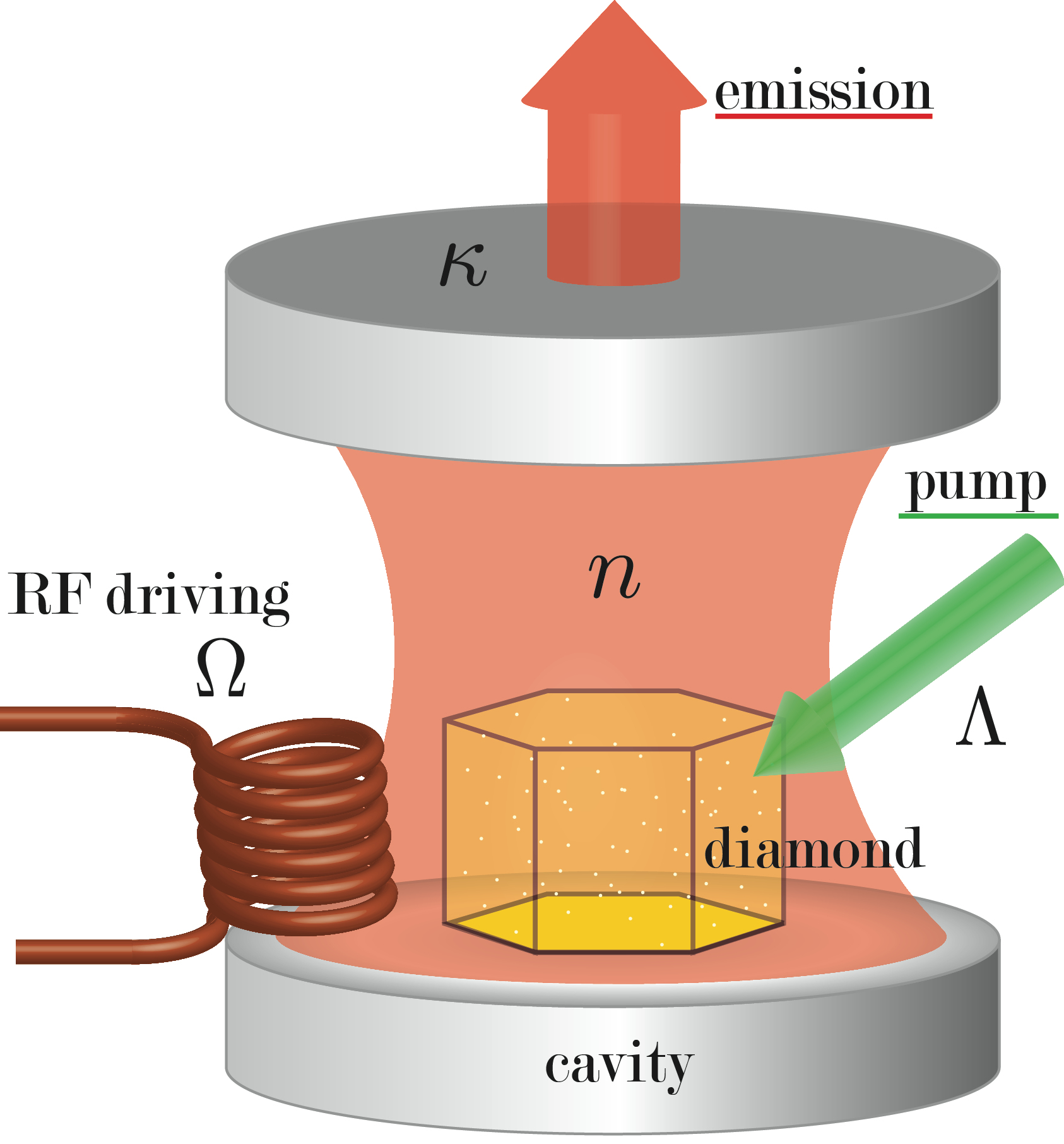}}
\hspace{1cm}
\subfloat[]{\includegraphics[scale=1]{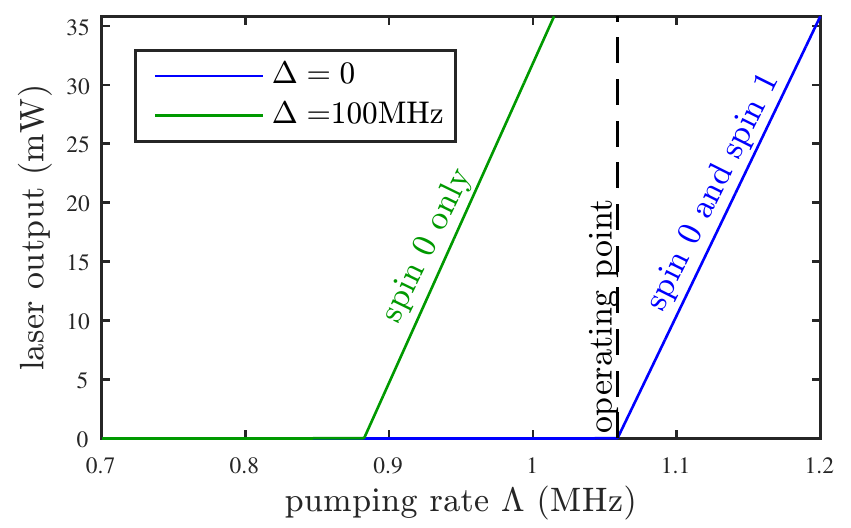}}\\
\subfloat[]{\includegraphics{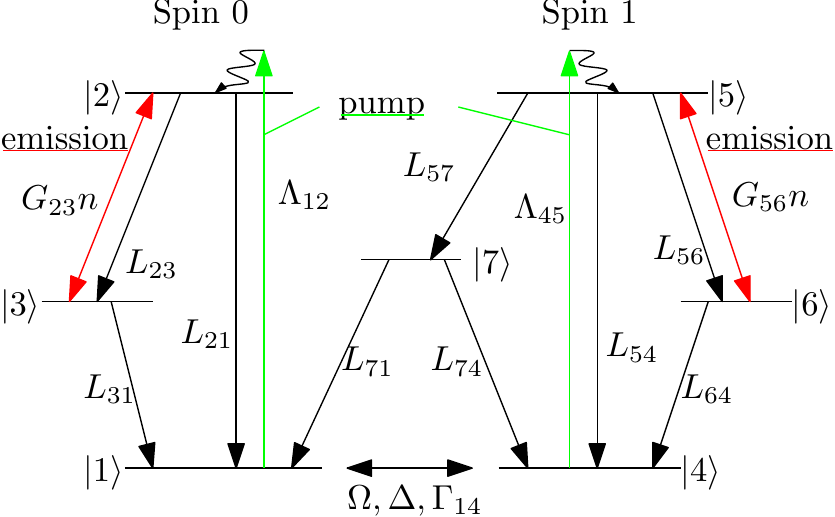}}
\caption{\label{fig concept}
(a)~Concept for a nitrogen-vacancy (NV$^-$) laser system using laser threshold magnetometry (LTM). NV$^-$ is the laser gain medium, pumped in the green and lasing on the red three--phonon sideband.  
(b)~The laser output over the pumping rate when the RF drive is off-resonant, $\Delta=100$MHz, or on-resonant, $\Delta=0$. The lasing threshold is dependent on the spin manifold. The operating point is chosen in between such that the laser turns on and off depending on the magnetic resonance of the spin manifolds and achieves maximal sensitivity. Setting the operating point to $\Lambda=1.06$ MHz ensures the laser turns off at $\Delta=0$ and maximises the laser output at $\Delta=100$ MHz. The Rabi frequency is $\Omega=3.67$ MHz. 
(c)~Reduced level structure for NV$^-$, breaking the system into the manifolds for spin 0 and spin 1 and highlighting the state transitions.  Mixing between the manifolds is only possible via the singlet pathway $L_{57}$, which takes population from the spin 1 manifold to the spin 0 manifold, and via the RF drive, which in the incoherent limit tends to equalise populations. The green pump laser lifts population into a phonon-added state just above $\ket{2}$ and $\ket{5}$, followed by a very rapid decay into $\ket{2},\ket{5}$. The $\ket{2} \leftrightarrow \ket{3}$ and $\ket{5} \leftrightarrow \ket{6}$ transitions emit into the cavity.
}
\end{figure*}

Here we propose an alternative mechanism for magnetometry, illustrated in Fig.~\ref{fig concept}(a). Crucially, instead of direct fluorescence measurements of NV$^-$ ensembles, a laser is constructed from the NV$^-$ centres themselves. This replaces the need for initialisation and projective measurement with an always-on laser signal output. The central idea is to use the change in fluorescence that arises because of the magnetic-field dependent spin populations, to shift the laser from below threshold to above threshold, see Fig.~\ref{fig concept}(b).  This gives rise to a coherent output signal from the NV$^-$ ensemble, indicative of the magnetic field strength, with multiple advantages.  The laser signal strength can vary by many orders of magnitude, greatly improving contrast, while direct fluorescence measurements require the discrimination between two relatively similar intensities (typically a $20\%$ difference in fluorescence levels for single NV$^-$ centres \cite{Maze2008}). Furthermore, stimulated emission creates photons in the same spatial mode, ideally permitting the collection of all such photons. In contrast, the collection efficiency of conventional NV$^-$ magnetometry is limited by the numerical aperture of the optics used to image the NV$^-$, which emit into all space. Finally the stimulated emission rate relative to the spontaneous emission rate changes with the magnetic field due to its dependence on the number of cavity photons. Thus the spin-dependent fluorescence difference is enhanced at the laser threshold by the competition between spontaneous and stimulated emission. Laser threshold magnetometry (LTM) leads to milliwatt photon power output and a commensurate increase in sensitivity reaching the fT/$\sqrt{\mathrm{Hz}}$ scale.

Diamond was investigated as a laser medium \cite{Rand1992} and used for Raman lasers \cite{Mildren2009}, diamond UV LEDs have been demonstrated \cite{Lohrmann2011}, and NV$^-$ centres have been coupled to cavities \cite{Faraon2011, Albrecht2013}. However, we are unaware of any diamond-based magnetometer with the sensitivities we predict here. LTM will lead to magnetometers with sensitivities 2-3 orders of magnitude better than existing NV$^-$ demonstrations and comparable to state-of-the-art SQUID and SERF magnetometers. 

\section{Results}
\subsection{A nitrogen-vacancy laser}
The electronic ground state manifold of the NV$^-$ centre is a spin one triplet, which at zero magnetic field has a spin-0 ground state, and nearly-degenerate spin $\pm 1$ states at 2.88~GHz, due to the crystal field splitting \cite{Loubser1978}. For our purposes we will treat it as a spin half system, assuming that optical pumping into the spin 0 ground state is perfect \footnote{In practice, the optical pumping into the spin 0 ground state is not perfect, with a flip probability of a few percent.  Inclusion of this imperfection did not significantly alter our calculations.}, and the radio-frequency (RF) field that induces spin flips is only resonant with one of the $\pm 1$ spin states.  We further assume that all of the processes except for the singlet pathway and the RF drive field are spin conserving.  Hence we can separate the problem into two manifolds, the spin 0 manifold and the spin 1 manifold. A schematic of the pertinent energy levels of the NV$^-$ system is given in Fig.~\ref{fig concept}(c). The $^3A_2$ spin 0 and spin 1 states are denoted by $\ket{1}$ and $\ket{4}$ respectively and have an energy difference that changes linearly with external magnetic field. An external RF drive can induce coherent Rabi oscillations at frequency $\Omega$ and detuning $\Delta$ between these two states, i.e.~$\Delta$ also changes linearly with external magnetic field. $\Gamma_{14}$ is the ground-state decoherence. A green pump laser drives population to a phonon-added state just above the $^3E$ spin 0 and spin 1 states, denoted by $\ket{2}$ and $\ket{5}$ respectively, followed by a rapid decay into $\ket{2},\ket{5}$. This effectively incoherent driving is spin-conserving: $\ket{1} \rightarrow \ket{2}$ and $\ket{4}\rightarrow \ket{5}$. The system decays from $\ket{2}, \ket{5}$ via several decay paths back to states $\ket{1}$ and $\ket{4}$: Direct decay creating a photon, indirect decay creating a photon of less energy and one or more phonons - the phonon sidebands, and decay via the singlet states $^1A_1$ and $^1E$, for simplicity represented by only one state $\ket{7}$. Due to this relatively long-lived singlet pathway a spin polarisation mechanism occurs such that spin~1 population is more likely to be pumped to spin~0. Because the singlet pathway is longer lived than the direct emission, the spin 0 state is slightly brighter than the spin~1 state. This mechanism is used elsewhere for both spin polarisation and readout of an individual NV$^-$ centre. We use the difference in fluorescence to shift the laser threshold based on external magnetic fields.

As the branching ratio to the 3 phonon sideband is the strongest \cite{Aharonovich2011, Rabeau2005, Doherty2013}, we denote the 3 phonon added ground state in the respective spin manifold by states $\ket{3},\ket{6}$ and consider lasing on the $\ket{2}\leftrightarrow\ket{3}$ and $\ket{5}\leftrightarrow\ket{6}$ transition while all other sidebands are included into the effective direct decays $L_{21}, L_{54}$. The phononic states $\ket{3},\ket{6}$ are only virtual levels and relax very rapidly via phononic processes to the ground states, $\ket{1},\ket{4}$.  Hence for our purposes we require population inversion between $\ket{3}$ and $\ket{2}$ or between $\ket{6}$ and $\ket{5}$ respectively. 
The $L_{ij}$ represent incoherent decay rates within the NV$^-$ centre and the $G_{ij}$ the coherent (cavity induced) transitions, $\Lambda_{12}, \Lambda_{45}$ represent effectively incoherent pumping rates from the green laser pump. Finally $n$ is the number of intracavity photons per NV$^-$ centre and $\kappa$ is the photonic loss rate from the cavity, which determine the red laser output power $P_{out}$:
\begin{align}
P_{out}=2\pi n N_{at} \kappa \hbar \nu_{23} 
\end{align}
where $N_{at}$ is the number of NV$^-$ centres in the cavity. The laser output power and pumping power both scale linearly with $\kappa$ in the relevant parameter regime (input via linear dependence on the operating point). This means laser pump power is another tunable parameter although in practice the output will be limited by the input power one can provide to the diamond crystal.

Optical charge state conversion can occur in NV$^-$ centres \cite{Fu2010, Waldherr2011, Beha2012, Chen2013, Aslam2013, Siyushev2013, Shields2015}, and the steady state NV$^-$/NV$^0$ ratio as a function of intensity has been observed in nanodiamond and single crystal samples\cite{Henderson2011, Chen2015}.  Because we are considering a large ensemble of NV$^-$ centres with effectively instantaneous charge conversion events, the steady state NV$^-$ population is the correct number of NV$^-$ centres to use in the laser threshold calculations.

\subsection{Realistic parameters}
For simplicity and due to lack of separate experimental data on each spin manifold, we will assume the cavity-induced transitions $G_{23}=G_{56}=G$ and all corresponding rates in the two spin manifolds to be the same $L_{23}=L_{56}, L_{31}=L_{64}$ and  $L_{21}=L_{54}$. The only exceptions are $L_{74}=(462$ns$)^{-1}$, $L_{71}=L_{74}/2$ and $L_{57}=(24.9$ns$)^{-1}$, based on ref.~\cite{Doherty2013}, which define the non-spin-conserving singlet decay path. 
The cavity-induced transitions $G$ describe the stimulated emission and absorption rates and are dependent on the cavity volume $V_c$, and the number of NV$^-$ centres $N_{at}$ 
\begin{align}
G=3 \nu_{23} L_{23} \lambda^3 N_{at} /(4 \pi^2 \Delta \nu_{23} V_c)
\end{align}
with the transition frequency $\nu_{23}=c/709$nm (1.75eV photon energy, see Fig.~11 in \cite{Doherty2013}), the corresponding wavelength $\lambda=709$nm$/2.4$, where 2.4 is the refractive index of diamond, the peak width of the three--phonon sideband $\Delta \nu_{23}\approx 24$THz, \footnote{(see figure 22 in \cite{Doherty2013} or figure 4 in \cite{Kehayias2013})}. Assuming a density of NV$^-$ centres of 1 per $(100$nm$)^3$, which corresponds to 5.7 ppb, a laser medium volume of 1mm$^3$ and an external cavity volume of $V_c=2$mm$^3$ we find a cavity-induced transition rate of $G_{23}=G_{56}=G=308$MHz. The spontaneous decay rate to the three-phonon sideband is $L_{23}=L_{56}=18$MHz and the direct decay rate $L_{21}=L_{54}=68.2$MHz \cite{Suphd, Su2008}. The phononic decay is much faster and set to $L_{31}=L_{64}=1$THz, since the exact value is unknown, but unimportant as it is fast compared to the other decay rates, and small compared with the transition frequencies. Since our scheme does not perform spin-echo of any kind the dephasing is set by the inhomogeneous coherence time $T_2^*$. For the assumed low density of 5.7 ppb NV$^-$ this time is determined by the $^{13}$C concentration, which can be reduced from the natural abundance of $\sim$1\% by isotopic purification \cite{Mizuochi2009}. We set the dephasing rate to $\Gamma_{14}=(1\mu$s$)^{-1}$ which is above already experimentally achieved values \cite{Mizuochi2009, Doherty2013}. The cavity loss rate is set to $\kappa=3$MHz, which corresponds to a cavity $Q=2 \pi \nu_{23}/\kappa=8.9\times 10^8$, neglecting all other losses. These values are easily achievable with an external cavity. This cavity design is large compared to those used in semiconductor laser diodes, but compatible with the dimensions of readily available synthetic diamond crystals.

\begin{figure*}
\subfloat[]{\includegraphics[scale=.88]{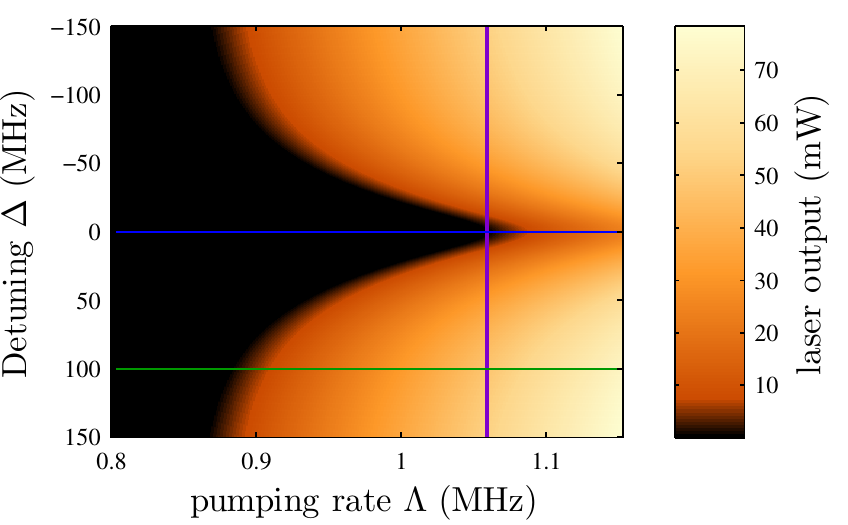}}
\subfloat[]{\includegraphics[scale=.88]{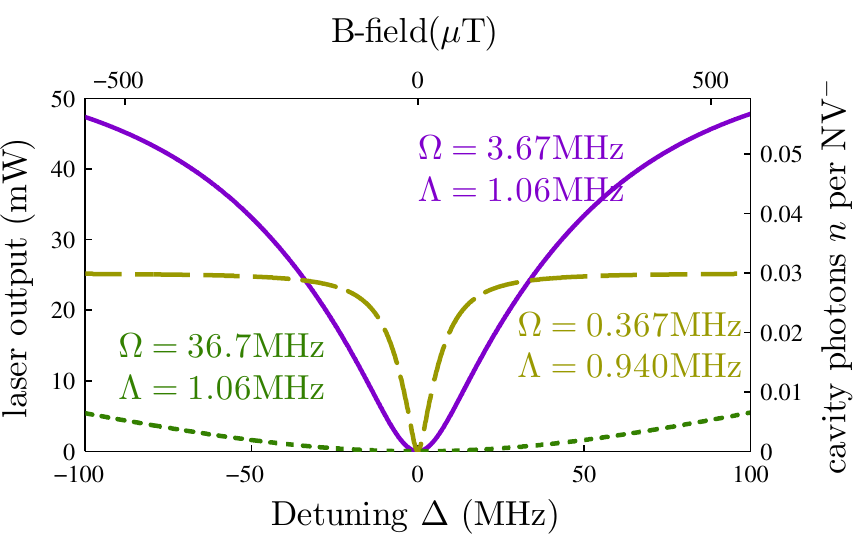}}\\
\caption{
(a)~Laser output power as a function of detuning and pumping rate for a fixed Rabi-frequency $\Omega=3.67$MHz. The green and blue horizontal lines correspond to the laser thresholds shown in Fig.~\ref{fig concept}(b) for the spin 0 and spin 0+1 cases respectively.  The operating point of the pumping rate is ideally set to $\Lambda=1.06$ MHz (purple line)
(b)~Laser output power as a function of detuning at the respective operating points. The detuning (bottom axis) and B-field (top axis) can be inferred from the laser output. The purple line (solid) corresponds exactly to the vertical line in (a). One can tune to higher precision (dashed) or greater measurement range (dotted) as required without changing the fabrication parameters: cavity loss rate $\kappa$, NV$^-$~density and coherence time. In this plot we have chosen $\kappa=3$~MHz, [NV$^-$]=5.7ppb, and $T_2^*=(\Gamma_{14})^{-1}=1~\mu$s.
\label{fig laser output}}
\end{figure*}

The NV$^-$ centres are aligned along 4 directions within the single crystal diamond. Our calculations assume the magnetic field to be oriented along one of them, and take into account the background fluorescence produced by the other three orientations. We note that even better sensitivities than our estimates here might be achieved by preferentially aligned NV$^-$ centres \cite{Edmonds2011} and polarisation-selective excitation\cite{Alegre2007}.

\subsection{Laser threshold magnetometry}
Figure \ref{fig concept}(b) shows the laser output power as a function of the laser pumping rate $\Lambda_{12}=\Lambda_{45}=\Lambda$ with resonant ($\Delta=0$) and off-resonant ($\Delta=100$MHz) RF drive. At off-resonant driving, the singlet pathway $L_{71}$ guarantees population to only be in the spin 0 manifold. At resonance, the RF drive mixes the populations of $\ket{1}$ and $\ket{4}$ and works against the singlet pathway, which has a long life-time relative to the triplet emission rates. At resonant driving, the singlet pathway behaves like an additional loss channel of the laser medium, which increases the lasing threshold. In other words, on resonance the emission is reduced due to some population being in the relatively long-lived, non-radiative state $\ket{7}$.

To operate as a high precision magnetometer we choose an operating point such that, on resonance (with population shared amongst the spin 0 and spin 1 manifold) the population is just below threshold, whereas off-resonance (when the system has been optically pumped into the spin 0 ground state) the system is above threshold. This operating range is around $\Lambda=1.06$MHz in Fig.~\ref{fig concept}(b) and \ref{fig laser output}(a). 

Figure \ref{fig laser output}(a) shows the laser ouput power $P_{out}$ in mW as a function of $\Delta$ and $\Lambda$. Once the pumping rate $\Lambda$ is set to the desired operating point (purple line), nonzero detuning pushes the system over the lasing threshold. The RF drive sets the initial value for $\Delta$, e.g. zero; any further changes to $\Delta$ are then caused by changes in the outer magnetic field, i.e.~any non-zero external magnetic field turns the laser on, with an intensity indicative of the magnetic field strength. With no detuning bias, the external magnetic field is given by
\begin{align}
B= \Delta \frac{\hbar}{g_e \mu_B} 
\end{align}
i.e.~$B/\Delta=5.68\;\mu$T/MHz, where $g_e$ is the Land\'{e} g-factor and $\mu_B$ the Bohr magneton.
In this way the device is a highly sensitive magnetic field sensor, or magnetometer, with very high contrast. 

Figure \ref{fig laser output}(b) shows how $P_{out}$ changes with $\Delta$ at the operating point $\Lambda=1.06$MHz (solid line). Selecting different values of the Rabi-frequency $\Omega$ and the pumping rate $\Lambda$ provides either increased dynamical measurement range (dotted) or increased measurement precision (dashed) as it increases the laser output gradient for small B-fields.  As both the RF drive and laser pump are externally controlled parameters, these adjustments do not require changes of fabrication parameters but simply represent different operating modes. The precision could furthermore be increased by scanning across the entire dip via a variation of the detuning to obtain one high-precision measurement of the magnetic field.
 
The device has a response time $t_r$ to sudden changes of the magnetic field which is essentially set by the slowest process out of RF driving $\Omega$, laser pumping rate $\Lambda$, cavity decay $\kappa$, and effective non-spin conserving transition rate. For the parameters of the purple solid line in Fig.~\ref{fig laser output}, the laser pumping rate  is the slowest process and sets $t_r\approx 1/\Lambda=0.94\mu$s (examples in Supplementary Material). This can be shortened by increasing $\kappa, \Lambda, \Omega$ such that the non-spin conserving transition rate sets $t_r\approx 1/L_{57} + 1/L_{71}=0.5\mu$s. Since this non-spin conserving transition rate is not precisely known experimentally, the device can also serve as a probe for this parameter.

\begin{figure*}
\subfloat[]{\includegraphics[scale=.88]{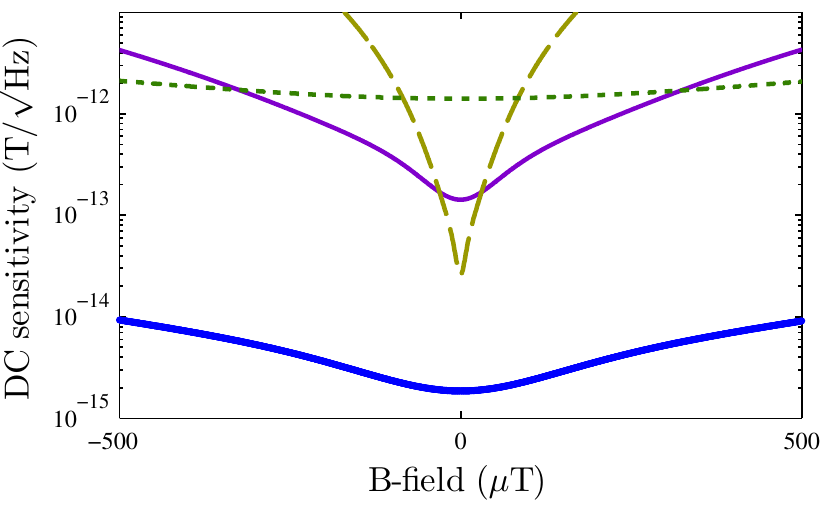}}\quad
\subfloat[]{\includegraphics[scale=.88]{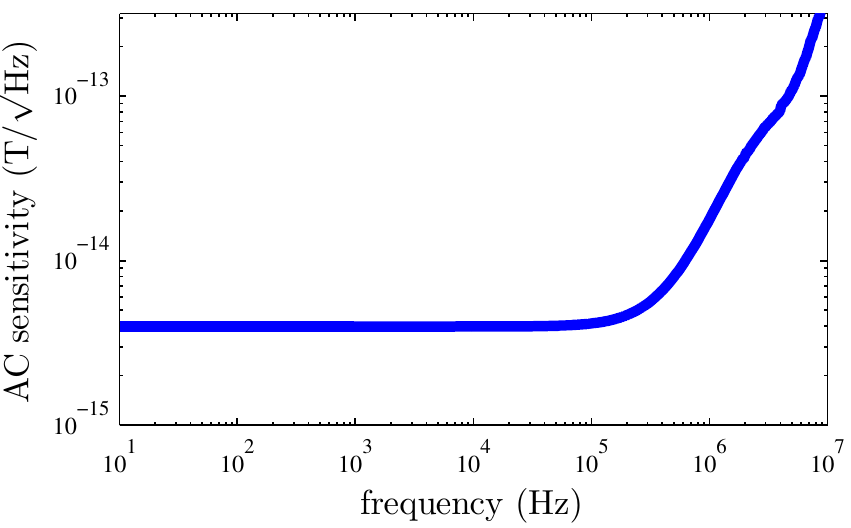}}
\caption{
(a)~D.C.~Sensitivity as a function of external magnetic field. Sensitivities down to 1.86 fT/$\sqrt{\rm{Hz}}$ (blue solid line) can be achieved by increasing the NV$^-$ concentration to 16ppm with a coherence time $T_2^*=0.181\mu$s, and choosing $\kappa=63.1$GHz, $\Lambda=10.4$MHz and $\Omega=6.14$MHz. The dashed, dotted and purple solid line are for the parameters as in figure 2(b). 
(b)~A.C.~Sensitivities as low as 3.97 fT/$\sqrt{\rm{Hz}}$ can be reached as a function of the external magnetic a.c.~field frequency. Device parameters are the same as for the blue solid line in (a). We apply an effective bias d.c. field of $B_o=164\mu$T to reach the region where the laser output has the strongest gradient with the external B-field.
}
\label{fig sensitivity}
\end{figure*}

\subsection{Sensitivity}
The shot-noise-limited sensitivity of the device is defined as $\eta \equiv B_{min}\sqrt{T}$ where $B_{min}$ is the minimum field that can still be detected in the total measurement time $T$. Since longer measurements in the presence of shot-noise decrease the error as $1/\sqrt{T}$ this factor must be multiplied with $B_{min}$ to achieve a $T$-independent value. The sensitivity is: 
\begin{align}
\eta_{dc}=\frac{dB}{dn} \sqrt{\frac{n}{N_{at} \kappa}} 
\end{align}
The sensitivities corresponding to the different tuning parameters of Fig.~\ref{fig laser output}(b) are plotted with corresponding colors and linestyle in Fig.~\ref{fig sensitivity}(a) as a function of the magnetic field strength being measured. 

Ultrahigh precision can be achieved by increasing our conservative estimates for the NV$^-$ density. However, for high NV$^-$ concentrations the $T_2^*$ time is reduced by the large number of N impurities and NV$^-$ to NV$^-$ interactions, which has negative effects on the sensitivity; diamond samples with high conversion efficiency from N to NV$^-$ and low numbers of other impurities are therefore ideal in maximising both NV$^-$ concentration and $T_2^*$ time. For the rest of the paper we use the combination of 16 ppm NV$^-$ concentration and $T_2^*=0.181 \mu$s, which has been measured in Ref.~\cite{Acosta2009}. 
These adaptations  enable an optimal sensitivity of $\eta=$1.86 fT$/\sqrt{\rm{Hz}}$, i.e.~a precision of 1.86 fT can be achieved in a 1-second-long measurement. This sensitivity is reached at the centre of the dip and behaves rather smoothly, see blue line in Fig.~\ref{fig sensitivity}(a). This is about 6 orders of magnitude better than several previous NV$^-$ magnetometry experiments \cite{Pham2011, Ishikawa2012, Fang2013, Jensen2014}, 2-3 orders of magnitude better than the best recent ensemble NV$^-$ experiments \cite{Clevenson2015, Wolf2014} and reaches the sensitivity of cryogenically operated SQUIDS. LTMs could therefore become a room-temperature operated alternative to SQUIDs. However compared to single NV$^-$ centre magnetometry we have sacrificed the nanometre spatial resolution since our device is about 1mm$^3$.

So far we have only considered the measurement of constant (d.c.) magnetic fields. If the sensor is placed in an oscillating (a.c.) magnetic field signal $B(t)=B_S \cos(\omega t)$ then the cavity photon number $n$ and the laser output will vary in phase with the signal. To achieve good response even for very small B-field amplitudes we measure in the presence of an effective d.c.~field bias to offset the laser output and maximise the response d$P_{out}/$d$B$ to external a.c.~fields. In the parameters of Fig.~\ref{fig sensitivity}(a) this offset is approximately 164 $\mu$T for the blue line. The a.c.~sensitivity $\eta_{ac}$ for a laser signal which is caused by an oscillating cavity photon number $n(t)=n_S \cos(\omega t) + n_o$ is calculated as:
\begin{align}
\eta_{ac}=\frac{dB_S}{dn_S} \sqrt{\frac{n_o I}{N_{at} \kappa}} 
\end{align}
where the factor $I=2.43$ stems from the signal's contribution in each oscillation period. The sensitivity as a function of frequency is shown in Fig.~\ref{fig sensitivity}(b). Sensitivities down to 3.97fT/$\sqrt{\rm{Hz}}$ are achieved for low frequencies and small signal amplitudes (here: $B_S = 1$ nT).  Sensitivities to frequencies above 0.1 MHz deteriorate with increasing frequency due to the finite response time $t_r\approx 0.5 \mu$s of the lasing process, which we identified above. 

\section{Discussion}
We have presented the concept of lasing threshold magnetometry (LTM) and shown that with NV$^-$ centres in diamond as the laser medium a sensitivity of 1.86 fT/$\sqrt{\rm{Hz}}$ can be achieved using a diamond volume of 1mm$^3$ with an NV$^-$ density of 16 ppm. This sensitivity is almost $10^3$ times better than current NV$^-$ demonstrations, and our magnetometer design is simpler and more robust as it does not require pulse sequences but operates on CW laser pump and RF drive. Our predicted sensitivity is equivalent to current state-of-the-art SQUID magnetometers.

The magnetometer can be kept relatively small and the laser output could be guided into an optic fibre, making the sensor very mobile. The device furthermore can be operated at room temperature. This is a significant technological advantage over the standard SQUID sensors, which need to be operated at cryogenic temperatures (below 10K). This could particularly improve Magneto-encephalography (MEG) \cite{Hamalainen1993, Proudfoot2014}, which measures the weak 10fT - 1pT magnetic fields \cite{Hamalainen1993} produced by brain activity with less distortion \cite{Proudfoot2014} than electro-encephalography (EEG).

We have chosen NV$^-$ centres as our laser medium as they have a one-directional non-spin conserving transition and its properties are very well studied. In principle, LTM can be performed with other color centres for example, the silicon-vacancy centre in diamond, which seems particularly promising \cite{Pingault2014, Rogers2014}

\section{Methods}
\subsection{Equations of motion and steady state solution}
We take as our reduced model a seven state system, with the intracavity photon field, shown in Fig.~\ref{fig concept}(a).  We also assume that the only non-negligible coherence $\rho_{14}$ is at the ground state transition.  The equations of motion for this simplified structure are:
\begin{align}
\dot{\rho}_{11} &= -2 \Omega \text{ Im}(\rho_{14}) - \Lambda_{12} \rho_{11} + L_{21} \rho_{22} + L_{31} \rho_{33} + L_{71} \rho_{77}, \nonumber \\
\dot{\rho}_{14} &= (i \Delta-\Gamma_{14}-\Lambda_{12}/2-\Lambda_{45}/2) \rho_{14} - i \Omega \left(\rho_{44} - \rho_{11}\right), \nonumber  \\
\dot{\rho}_{22} &= \Lambda_{12} \rho_{11}- \left(L_{21} + L_{23} \right)\rho_{22} - G_{23} \left(\rho_{22} -\rho_{33} \right)n, \nonumber \\
\dot{\rho}_{33} &= L_{23} \rho_{22} - L_{31} \rho_{33} - G_{23} \left(\rho_{33} - \rho_{22}\right) n, \nonumber \\
\dot{\rho}_{44} & = 2 \Omega \text{ Im} (\rho_{14}) - \Lambda_{45} \rho_{44} + L_{54} \rho_{55} + L_{64} \rho_{66} + L_{74} \rho_{77}, \nonumber \\
\dot{\rho}_{55} &= \Lambda_{45} \rho_{44} - \left(L_{54} + L_{56} + L_{57}\right)\rho_{55} - G_{56}\left(\rho_{55} - \rho_{66}\right) n, \nonumber \\
\dot{\rho}_{66} &= L_{56}\rho_{55} - L_{64}\rho_{66} - G_{56}\left(\rho_{66} - \rho_{55}\right) n, \nonumber \\
\dot{\rho}_{77} &= L_{57} \rho_{55} - \left(L_{71} + L_{74}\right) \rho_{77}, \nonumber \\
\dot{n} &= G_{23} \left(\rho_{22} - \rho_{33}\right) n  + G_{56} \left(\rho_{55} - \rho_{66}\right) n - \kappa n, \label{eq:PhotNum}
\end{align}
where the $\Lambda_{ij}$ represent effectively incoherent pumping rates from the green laser pump, which model the coherent excitation to a phonon-added state just above $\ket{2}$, followed by a rapid decay into $\ket{2}$. The $L_{ij}$ represent incoherent decay rates within the NV$^-$ centre and the $G_{ij}$ the coherent (cavity induced) transitions. The strength of the RF drive is given by $\Omega$ - the Rabi frequency, which has detuning $\Delta$, whilst $\Gamma_{14}$ is the ground-state decoherence.  The $\rho$ represent the density matrix elements for the NV$^-$~centres (fraction of the population in each state for the on-diagonal elements).  Finally $n$ is the number of intracavity photons per NV$^-$ centre, $\kappa$ is the photonic loss rate from the cavity and conservation of population implies $\sum_{i} \rho_{ii} = 1$.

The steady state solution of the equations of motion gives insight into the lasing threshold and laser output of the system and their dependence on the tunable parameters, such as laser pumping rate $\Lambda_{12}, \Lambda_{56}$, Rabi frequency $\Omega$ and detuning $\Delta$. We obtain the steady state by diagonalisation of the superoperator for the density matrix and subsequent solution of equation \ref{eq:PhotNum}, see supplementary material. We have a full analytical solution and inserted the realistic parameters given in the text. 

\begin{figure*}
\subfloat[]{\includegraphics[scale=.88]{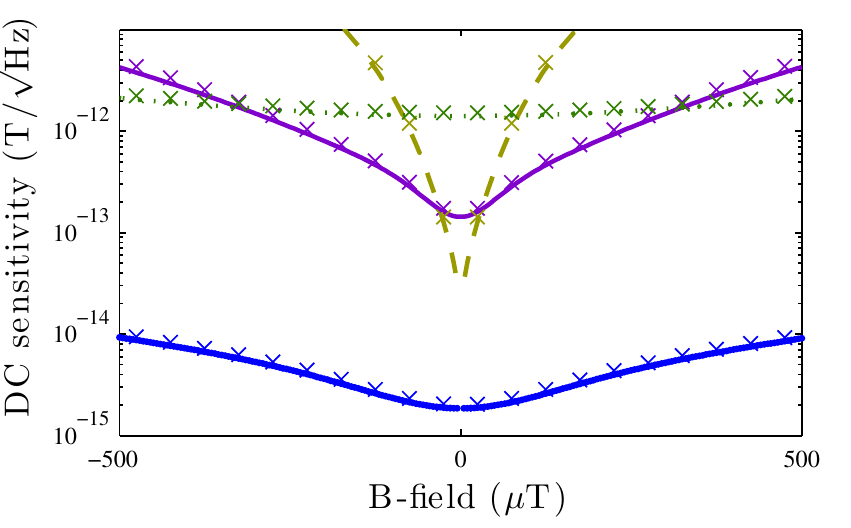}}
\subfloat[]{\includegraphics[scale=.88]{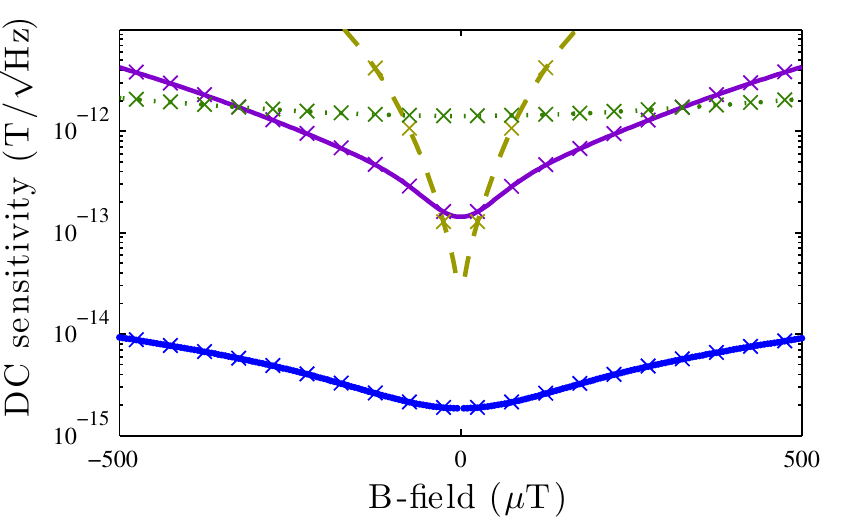}}
\caption{(a) LINES: A reproduction of Fig.~\ref{fig sensitivity}(a) where $L_{27}=0$.   CROSSES: Same calculation with an additional transition rate $L_{27}=0.1 L_{57}$. The resulting changes in the sensitivity values are below 10\%. This shows that the results do not critically depend on the rate $L_{27}$ being exactly zero.    (b) Same as in Fig.~\ref{fig both spin transitions}(a) but with a reduced $L_{27} = 0.01 L_{57}$.  Here the differences between the lines and the crosses are negligible.}
\label{fig both spin transitions}
\end{figure*}

In all calculations we ignored the possibility of a smaller transition rate in the other spin direction, from $\ket{2}$ to $\ket{7}$. Such a non-radiative rate has been found to be negligible, relative to the radiative decay \cite{Doherty2013}, and
``radiative non spin-conserving transitions ... are less than 1\% of the spin-conserving radiative decay rate'' \cite{Doherty2013}. An inclusion of such a rate  in the calculations, even at a higher non-negligible value of $L_{27}=0.1 \, L_{57}$, yielded only minor changes in the results, particularly the changes in the sensitivity values stay below 10\%, see Fig.~\ref{fig both spin transitions}.

Similar to other lasing schemes, our LTM requires a stable laser pump rate $\Lambda$ since fluctuations would lead to corresponding changes in the laser output, which look like signal changes. A stable source and/or laser amplitude stabilizers can help to minimize fluctuations. Furthermore the remaining input fluctuations can be monitored, by insertion of a beam splitter into the laser pump beam, and removed from the output signal. 

\ack
We acknowledge useful conversations with W.~J.~Munro, B.~Moran, B.~C.~Gibson, M.~Doherty, P.~Wilksch and N.~Vogt. ADG also acknowledges the ARC for financial support (DP130104381). 


\bibliographystyle{unsrt}
\bibliography{aaaPublication}

\end{document}